\documentclass[manuscript]{aastex}

\usepackage{url}
\usepackage{multirow}
\usepackage{graphicx}

\makeatletter

\newcommand{\Rmnum}[1]{\expandafter\@slowromancap\romannumeral #1@}
\makeatother

\begin{document}

\title{\emph{HST} ROTATIONAL SPECTRAL MAPPING OF TWO L-TYPE BROWN DWARFS:
VARIABILITY IN AND OUT OF WATER BANDS INDICATES HIGH-ALTITUDE HAZE LAYERS}


\author{Hao Yang}
\affil{Department of Astronomy, University of Arizona, 933 N. Cherry Avenue, Tucson, AZ 85721, USA} 
\email{haoyang@email.arizona.edu}

\author{D\'aniel Apai}
\affil{Department of Astronomy, University of Arizona, 933 N. Cherry Avenue, Tucson, AZ 85721, USA;} 
\affil{and Department of Planetary Sciences, 1629 E. University Blvd, Tucson, AZ 85721, USA}

\author{Mark S. Marley}
\affil{NASA Ames Research Center, Naval Air Station, Moffett Field, Mountain View, CA 94035, USA} 

\author{Didier Saumon}
\affil{Los Alamos National Lab, Los Alamos, NM 87545, USA} 

\author{Caroline V. Morley}
\affil{Department of Astronomy and Astrophysics, University of California, Santa Cruz, CA 95064, USA} 

\author{Esther Buenzli}
\affil{Max-Planck-Institut f\"ur Astronomie, K\"onigstuhl 17, 69117 Heidelberg, Germany} 

\author{\'Etienne Artigau}
\affil{D\'epartement de Physique, Universit\'e de Montr\'eal, C.P. 6128 Succ. Centre-ville, Montr\'eal, QC H3C 3J7, Canada} 

\author{Jacqueline Radigan}
\affil{Space Telescope Science Institute, 3700 San Martin Drive,Baltimore, MD 21218, USA} 

\author{Stanimir Metchev}
\affil{Department of Physics and Astronomy, Western University, 1151 Richmond Street, London, ON N6A 3K7, Canada }

\author{Adam J. Burgasser}
\affil{Center for Astrophysics and Space Science, University of California San Diego, La Jolla, CA 92093, USA}

\author{Subhanjoy Mohanty}
\affil{Imperial College London, 1010 Blackett Lab, Prince Consort Road, London SW7 2AZ, UK}

\author{Patrick J. Lowrance}
\affil{Infrared Processing and Analysis Center, MS 100-22, California Institute of Technology, Pasadena, CA 91125, USA} 

\author{Adam P. Showman}
\affil{Department of Planetary Sciences, University of Arizona, 1629 University Boulevard, Tucson, AZ 85721, USA} 

\author{Theodora Karalidi}
\affil{Department of Astronomy, University of Arizona, 933 N. Cherry Avenue, Tucson, AZ 85721, USA} 

\author{Davin Flateau}
\affil{Department of Planetary Sciences, 1629 E. University Blvd, Tucson, AZ 85721, USA}

\author{Aren N. Heinze}
\affil{Department of Physics and Astronomy, State University of New York, Stony Brook, NY 11794-3800, USA}


\begin{abstract}
We present time-resolved near-infrared spectroscopy of two L5 dwarfs, 
2MASS J18212815+1414010 and 2MASS J15074759-1627386,
observed with the Wide Field Camera 3 instrument on the \emph{Hubble Space Telescope} (\emph{HST}). 
We study the wavelength dependence of rotation-modulated flux variations between 1.1 $\mu$m and 1.7 $\mu$m. 
We find that the water absorption bands 
of the two L5 dwarfs at 1.15 $\mu$m and 1.4 $\mu$m vary at similar amplitudes as the adjacent continuum.
This differs from the results of previous \emph{HST} observations of L/T transition dwarfs, 
in which the water absorption at 1.4 $\mu$m displays variations of about half of the amplitude at other wavelengths. 
We find that the relative amplitude of flux variability out of the water band with respect to that in the water band
shows a increasing trend from the L5 dwarfs toward the early T dwarfs.
We utilize the models of \citet{saumon2008} and find that the observed 
variability of the L5 dwarfs can be explained by the presence of spatially varying high-altitude haze layers 
above the condensate clouds. Therefore, our observations 
show that the heterogeneity of haze layers - the driver of the variability - must be located at very low 
pressures, where even the water opacity is negligible. In the near future, the rotational spectral mapping 
technique could be utilized for other atomic and molecular species to probe different pressure levels 
in the atmospheres of brown dwarfs and exoplanets and uncover both horizontal and vertical cloud structures.




\end{abstract}


\section{INTRODUCTION}

Since the recent discovery of three early T dwarfs with high-amplitude flux variability 
as large as $26$\% in the near-infrared (near-IR) \citep{artigau2009, radigan2012, biller2013}, 
ground- and space-based observations have revealed that low-level variabilities are common 
for brown dwarfs of diverse spectral types across a wide range of wavelengths
(e.g., \citealp{heinze2013}; \citealp{gillon2013};  \citealp{buenzli2014a}; \citealp{burgasser2014};  
\citealp{wilson2014}; \citealp{radigan2014a}; \citealp{radigan2014b}; Metchev et al. 2014, submitted).

Condensate clouds are believed to play a major role in the atmospheres of brown dwarfs 
\citep[e.g.,][]{tsuji1996,jones1997,burrows2000,allard2001,ackerman2001,tsuji2002,helling2008,saumon2008, stephens2009},
and heterogeneous cloud covers combined with fast rotation are thought to produce the observed 
flux variability.  For the L dwarfs, silicate cloud layers form from condensation and 
strongly impact the emergent flux at these effective temperatures \citep[e.g.,][]{chabrier2000,lodders2006}. 
Mid and late T dwarfs are regarded as generally cloud-free objects due to clouds dispersing or 
sinking below their photospheres \citep[e.g.][]{ackerman2001,burgasser2002}, though sulfide clouds might still 
exist at altitudes high enough to affect the atmosphere \citep{morley2012}. 
As transtional objects between L dwarfs and mid/late T dwarfs, early T dwarfs show intermediate 
near-IR colors. Based on an emerging number of variable brown dwarfs, 
these objects likely have heterogeneous cloud coverage \citep[e.g.,][]{apai2013,radigan2014a,crossfield2014,buenzli2014a,buenzli2014b}, 
caused by varying cloud thickness, i.e., thin-thick clouds, but no cloud holes \citep{radigan2012, apai2013}.
Recent models have also shown that temperature perturbations, potentially 
arising due to atmospheric circulation, could also induce periodic and aperiodic flux variations 
\citep{showman2013, zhang2014, robinson2014, morley2014}.


Studies of clouds are not only important in understanding brown dwarf atmospheres, 
but also can shed light on the atmospheric properties of exoplanets \citep{kostov2013}.
Recent observations of \emph{transiting exoplanets} show that high-altitude clouds or haze layers 
may exist at pressure levels of 1 mbar or even lower 
\citep[e.g.,][]{sing2011, kreidberg2014, knutson2014}. While more detailed characterization 
of the exoplanetary atmospheres is limited by current instrumention,
studies of clouds or haze layers in brown dwarf atmospheres offer essential insights for reference.


Rotational spectral mapping is a powerful technique for uncovering cloud structures on brown dwarfs.  
\citet{buenzli2012} utilized the unique capabilities of the Wide Field Camera 3 (WFC3) instrument on 
the \emph{Hubble Space Telescope} (\emph{HST}) and detected phase shifts among light curves of 
different wavelength bands in the T6 dwarf 2M22282889-4310262. 
The multi-layer rotational maps revealed heterogeneous atmospheric structures in both horizontal 
and vertical directions for the first time. \citet{apai2013} analyzed 
high-precision time-resolved WFC3 spectra of two L/T transition dwarfs, 2MASS J21392676+0220226 
(hereafter, 2M2139) and 2MASS J01365662+0933473 \citep[hereafter SIMP0136,][]{artigau2006}. 
Modeling the spectral and color changes showed that explaining the brightness variations requires 
a combination of thick and thin clouds. \citet{burgasser2014} monitored WISE J104915.57-531906.1 
\citep[or Luhman 16,][]{luhman2013} both photometrically and spectroscopically, and were able 
to reproduce the observed spectral variability with a brightness temperature two-spot model.
By analyzing \emph{HST}/WFC3 spectral time series, \citet{buenzli2014b} found that Luhman 16B varies
at all wavelengths from 1.1 to 1.6 $\mu$m with amplitudes ranging from 7\% to 11\%, and a two-component
thin-thick cloud model could explain most of the variability.
So far, this technique has been applied to mostly T dwarfs and only very short ($\sim$ 40 min) 
time series of a few L dwarfs without covering a full rotation in an \emph{HST} snapshot survey 
\citep{buenzli2014a} are available. Thus, there is currently limited observational information 
of time-variable components of the mean atmospheric structure of L dwarfs.

In this letter, we report \emph{HST}/WFC3 spectral mapping of two L5 dwarfs, 
2MASS J18212815+1414010 and 2MASS J15074769-1627386 (here after, 2M1821 and 2M1507). 
2M1821 was discovered by \citet{looper2008} and exhibits red near-IR colors ($J$-$K$ = 1.78) 
and silicate absorption at $9-11$~$\mu$m \citep{cushing2006}, which they concluded was indicative of 
unusually large dust opacity in the atmosphere, possibly due to low surface gravity or high metallicity. 
\citet{gagne2014} identified 2M1821 as a young field dwarf that shows signs of low surface gravity. 
2M1507 was discovered by \citep{reid2000} and also shows weak $9-11$~$\mu$m silicate
absorption \citep{cushing2006}.
Both L5 dwarfs have been observed to 
be variable in \emph{Spitzer} IRAC channels 1 and 2 (Metchev et al. 2014, submitted).
We study the wavelength dependence of their near-IR variabilities, and compare the results with
early T dwarfs, 2M2139, SIMP0136, and Luhman 16B.
In \S2, we describe our observations and the data reduction process. In \S3, we present 
the spectral variation of the two L dwarfs, which is followed by a discussion of the 
results in \S4. We summarize our results in \S5.


\section{OBSERVATIONS AND DATA REDUCTION}
The \emph{Spitzer Space Telescope} Cycle-9 Exploration Science Program, \emph{Extrasolar Storms} 
(PI: D. Apai), uses coordinated multi-epoch \emph{HST} and \emph{Spitzer} rotational phase maps 
of six brown dwarfs to characterize cloud evolution and dynamics of ultracool atmospheres over 
a large range of timescales. The observations presented here are part of the coordinated \emph{HST} 
component of the \emph{Extrasolar Storms} program. We obtained near-IR spectra of 2M1821, 2M1507, 
and the T2 dwarf SIMP0136 with the WFC3 G141 grism. 

Each target in our program was observed over 3 or 4 consecutive orbits each in two separate visits.
During each orbit, a direct image was first obtained through the F132N filter for wavelength 
calibration, followed by a number of dispersed images with the G141 grism. To avoid detector buffer 
dumps and maximize observing time in each orbit, subarrays of 256$\times$256 pixels on the detector 
were used, corresponding to a field of view of $\sim$ 30$\arcsec\times$30\arcsec. The spectra were kept on 
the same pixels for all exposures so that systematic errors caused by pixel-to-pixel sensitivity 
variations are avoided. The observations are summarized in Table \ref{obs}.

For data reduction, we downloaded spectral images processed by the standard WFC3 pipeline from the MAST 
archive\footnote{\url{http://archive.stsci.edu}}, and then utilized custom IDL routines and 
the PyRAF software package aXe\footnote{\url{http://axe-info.stsci.edu}} to extract the slitless spectra. 
The detailed data reduction process is described in \citet{apai2013} and \citet{buenzli2014a}. Briefly, 
in the two-dimensional spectral images (.flt files) already processed by the WFC3 pipeline, we first 
corrected cosmic rays and bad pixels flagged by the pipeline. Then we embedded the subarray images 
into full-frame ones so that aXe can use full-frame standard instrument calibration images. The \emph{axeprep}
routine was used to subtract sky background before the \emph{axecore} routine was applied to extract the spectra 
with a fixed 8-pixel extraction window. The reduced G141 grism spectra provide a wavelength coverage of 
1.05--1.7 $\mu$m and a spectral resolution of $\sim$ 130. The uncertainty level (including photon noise, 
readout and background noise) for our observations is about 0.3\% and is estimated using the observed 
spectra and the WFC3 IR Spectroscopic Exposure Time Calculator (version: 22.1.2).

Archival \emph{HST}/WFC3 observations of SIMP0136 and the T2.5 dwarf 2M2139 from GO Program 12314 
(PI: D. Apai) were also downloaded for comparison purposes and reduced in the same fashion described above. 

\begin{figure}[ht]
  \begin{center}
    \includegraphics[scale=0.6]{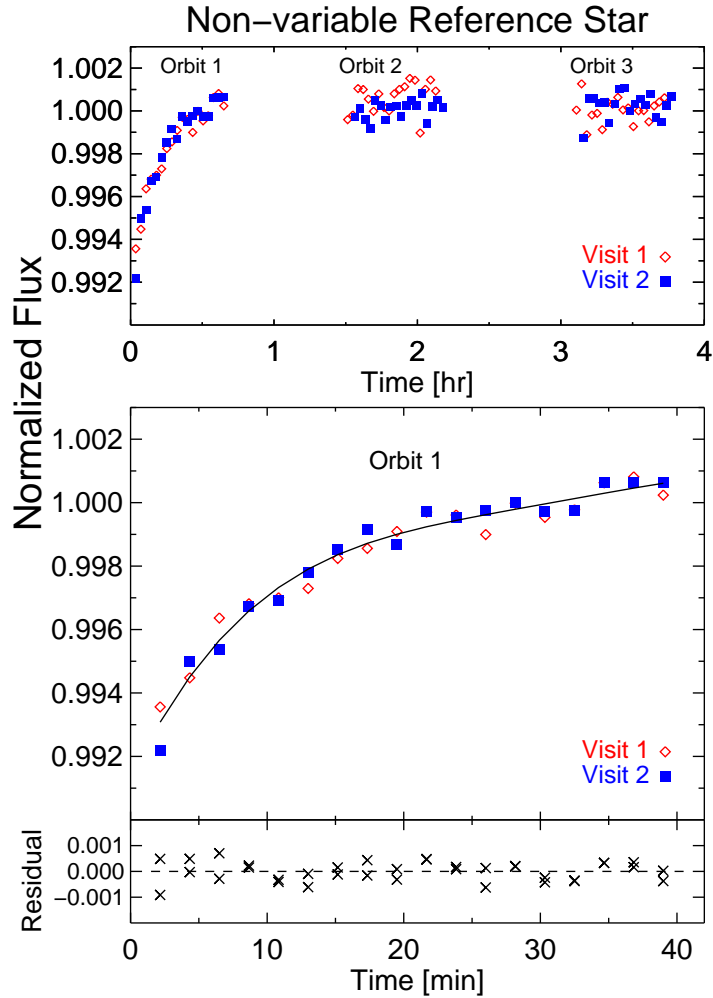}
       \caption{Correction of the ramp based on $J$-band light curves of a non-variable reference star. 
          \emph{Top}: $J$-band light curves of the reference stars from two \emph{HST} visits. Each visit
          has three orbits of observations. The first orbits show an increasing ramp, while the second and third 
          orbits have no substantial flux variation above the uncertainty level.
          \emph{Bottom}: the ramp in the first orbits is fitted with a fourth-order polynomial function, and the residuals after
          applying the polynomial correction have a standard deviation of 0.00038. }
           \label{ramp}
              \end{center}
    \end{figure}

During the first orbit in each visit, there was a common steep increase in brightness due to a systematic 
ramp effect. This is exemplified by the $J$-band light curve of a non-variable reference star in the field 
of view of 2M1821 (Figure~\ref{ramp}). The $J$-band fluxes are calculated by integrating each spectrum 
convolved with the 2MASS $J$-band spectral response curve \citep{cohen2003}. The first orbit shows a flux 
increase of nearly 1\%, while the second and third orbits shown stable flux levels within uncertainty 
($\sim$0.3\%). As discussed in \citet{apai2013} and \citet{buenzli2014a}, the ramp is found to be largely 
independent of wavelength and of object brightness. To remove the ramp effect, we fit a fourth-order 
polynomial function to the light curves of the first orbits for the non-variable reference star 
(bottom panels of Figure~\ref{ramp}), and applied the correction to the first-orbit observations of 
the variable brown dwarfs.



\section{RESULTS}

\begin{figure}[ht]
  \begin{center}
    \includegraphics[scale=0.65]{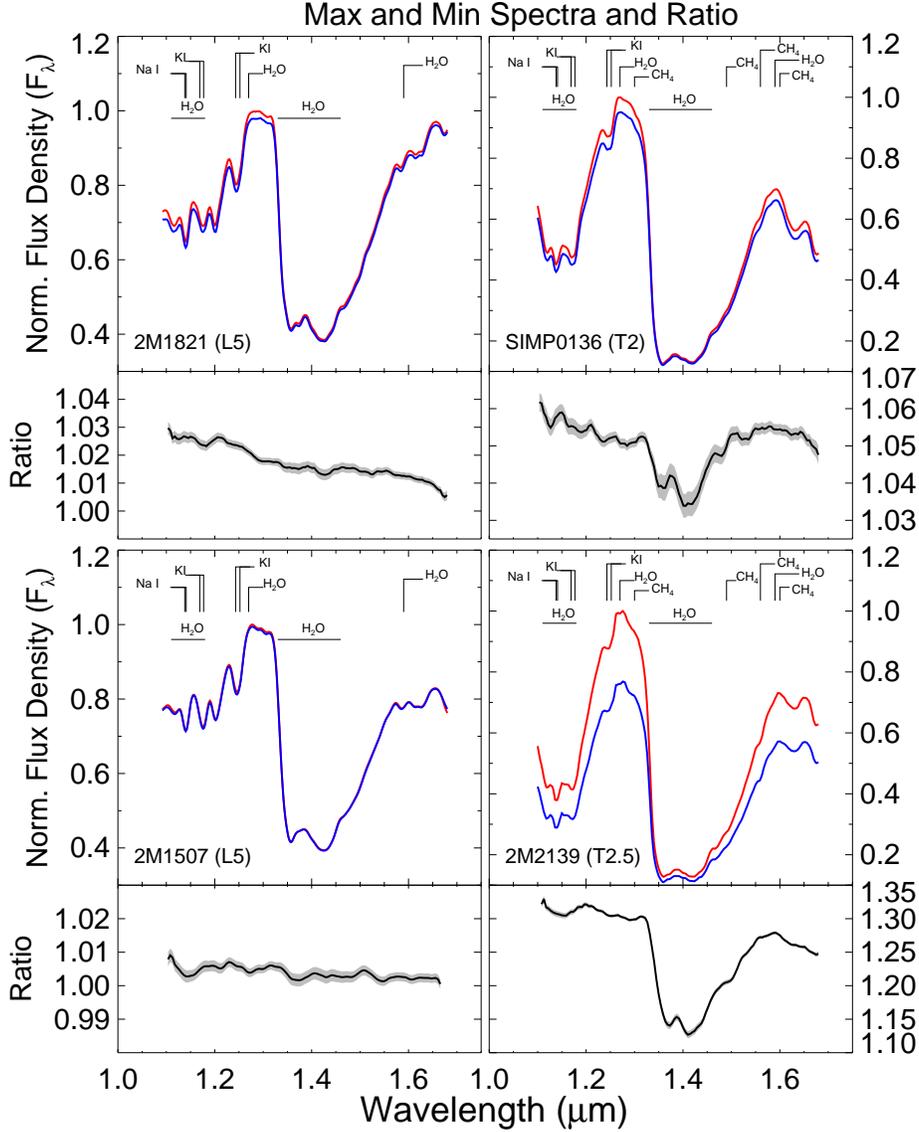}
       \caption{ Brightest (\emph{red}) and faintest (\emph{blue}) 
                 spectra and their ratio (\emph{minor panels}) during one \emph{HST} visit for 
                 2M1821, 2M1507, SIMP0136, and 2M2139, respectively. 
                 To increase the signal-to-noise, the brightest/faintest spectrum is from 
                 median combining the six brightest/faintest spectra from the same orbit. 
          The gray band marks the uncertainty level in the ratio of the brightest and faintest spectra.}
           \label{maxminspectra2}
              \end{center}
    \end{figure}

Our spectral time series of 2M1821 and 2M1507 reveal brightness variations between 1.1 $\mu$m and 1.7 $\mu$m.
In Figure~\ref{maxminspectra2}, we show their brightest and faintest spectra from
respective \emph{HST} Visit 1. Also shown for comparison are the spectra of the T2 dwarf SIMP0136 
and re-reduced archival data of the T2.5 dwarf 2M2139. The spectra of both the L5 dwarfs and early T dwarfs 
exhibit prominent absorption features of alkali elements  
and water. The L5 dwarfs have stronger \ion{Na}{1} and \ion{K}{1} absorption lines, while the early T dwarfs 
show deeper water absorption bands near 1.15 and 1.4 $\mu$m regions along with methane absorption features.


We compare the ratios of the brightest and faintest spectra in an \emph{HST} visit and discover that 
the variation of water-band absorption around 1.4 $\mu$m behaves differently for the two L5 dwarfs 
and the two L/T transition dwarfs. \citet{apai2013} first discovered for 
SIMP0136 and 2M2139 that the water band around 1.4 $\mu$m varies at a reduced amplitude compared to 
the continuum and other atomic and molecular absorption features. The same reduced water-band variability
is also found for Luhman 16B \citep{buenzli2014b}. However, for the two L5 dwarfs,
we find that the ratio of the brightest over faintest spectra shows generally weak wavelength dependence 
between 1.1 and 1.7 $\mu$m, and the water band around 1.4 $\mu$m varies at similar amplitudes as the 
adjacent continuum. 

\begin{figure}[ht]
  \begin{center}
    \includegraphics[scale=0.65,angle=90]{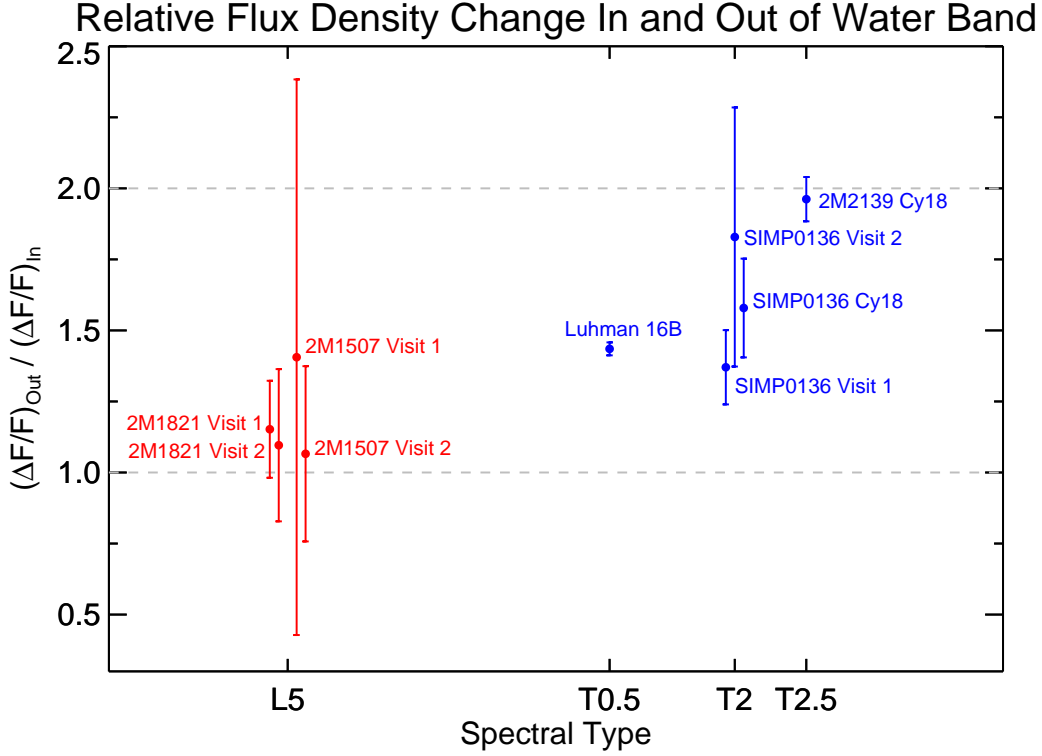}
       \caption{ Ratio of relative flux density changes in and out of the water band between the brightest and faintest spectra 
                 for the two L5 dwarfs (\emph{red}) and three L/T transition dwarfs (\emph{blue}).
                 The relative flux density change in the water band, $(\Delta F/F)_{\rm{In}}$, is the ratio of averaged flux densities
                 over the WFC3 F139M bandpass, and the relative flux density change out of the water band, $(\Delta F/F)_{\rm{Out}}$,
                 is the ratio of averaged flux densities over both the WFC3 F127M and F153M bandpasses. 
                 The gray dashed lines mark where $(\Delta F/F)_{\rm{Out}}$ is the same as $(\Delta F/F)_{\rm{In}}$ and 
                 twice the value of $(\Delta F/F)_{\rm{In}}$, respectively.
                 The values for Luhman 16B are calculated from data published in \citet{buenzli2014b}.
                 The relative flux variations out of the water band with respect to that in the water band 
                 shows an increasing trend from L5 dwarfs toward early T dwarfs.}

           \label{fluxchange}
              \end{center}
    \end{figure}

To further illustrate the different behavior in the variability in and out of the 1.4 $\mu$m water band, 
we perform synthetic photometry to measure the average flux density changes between the brightest 
and faintest spectra in several WFC3 medium bandpasses. We use the F139M filter to capture the average flux 
density in the water-absorption band and calculate the relative flux density change between the brightest
and the faintest spectra, $(\Delta F/F)_{\rm{In}}$.
Similarly, we measure the flux density averaged over the F127M and F153M filters and calculate the relative variation
in the average flux density out of the water-absorption band between the brightest and faintest spectra, $(\Delta F/F)_{\rm{Out}}$. 
Then we take the ratio between the relative flux density changes in and out of the water band, 
$(\Delta F/F)_{\rm{Out}} / (\Delta F/F)_{\rm{In}}$.
During the \emph{HST} Visit 1 of 2M1821, e.g., the relative change in average flux density is 
1.77\% $\pm$ 0.11\%  out of the water band and 1.54\% $\pm$ 0.21\% in the water band, and the ratio of the two
is 1.15 $\pm$ 0.17.
As shown in Figure~\ref{fluxchange}, the ratio $(\Delta F/F)_{\rm{Out}} / (\Delta F/F)_{\rm{In}}$ displays an increasing trend
from the L5 dwarfs to the early T dwarfs. The L5 dwarfs show similar relative flux variation in and out of the water band, 
while the L/T dwarfs have greater relative flux change out of the water band.
Such a trend with spectral types remains in observations of different epochs, even though the relative amplitudes 
of flux variation are different from visit to visit. The time between the two \emph{HST} visits for targets
in the Extrasolar Storms program is between one to three weeks, and for SIMP0136, the observations from two \emph{HST} cycles 
are separated by two years.

\section{DISCUSSION}

We investigate the near-IR spectral variability of the L5 dwarfs 2M1821 and 2M1507, 
and we find that the variations of the water-band absorption at 1.4 $\mu$m exhibits pronounced 
differences between the two L5 dwarfs and two L/T transition dwarfs.
We propose that such different behaviors could be due to the difference in the height of 
the dust particles in the atmospheres. 

We propose a toy model that can quantitatively explain the observed behavior of the L5 and the L/T dwarfs.
We assume that the intensity modulations, $\Delta I_{\rm{int}}$, are introduced at an altitude, $z$, and that 
the dust layer causing the modulations is not emitting.
The optical depth at $z$ measured from the top of the atmosphere is greater in the water band than in the adjacent continuum 
($\tau_{\rm{water}} > \tau_{\rm{cont}}$), but both are of the order of one. At a specific wavelength, $\lambda$, the modulations seen by the observer 
follow the Beer-Lambert law:
$\Delta I_{\rm{obs}} =  \Delta I_{\rm{int}} \cdot e^{- \tau_{\lambda}}$.

Then the relative variation in and out of the water band will be:

$$\epsilon = {  \Delta I_{\rm{obs, water}} \over \Delta I_{\rm{obs, cont}}   } = { \Delta I_{\rm{int}} \cdot e^{- \tau_{\rm{water}} } \over \Delta I_{\rm{int}} \cdot e^{-\tau_{\rm{cont}}}  } $$

or $\epsilon = e^{-(\tau_{\rm{water}} - \tau_{\rm{cont}})} $.

In this scenario, if the modulations are introduced high in the atmosphere, 
the optical depth difference between in 
and out of water band will be negligible, leading to $\epsilon \sim$ 1, as observed in the L5 dwarfs.
If, however, the modulations are introduced deeper, the optical depth difference will be more significant, 
leading to reduced variability amplitude in the water band, as observed in L/T dwarfs.
This simplistic model provides a correct relative 
variability amplitude difference between the continuum and water bands for both the L5 and the L/T cases.

When \citet{looper2008} discovered 2M1821, they found several lines of evidence indicating an unusually dusty 
atmosphere, including unusually red slopes throughout the $Z$, $Y$, and $J$ band, red near-IR colors, 
weak H$_2$O absorption, and silicate absorption at $9-11$~$\mu$m.  They considered either high clouds or high 
metallicity as the explanation for the high dust opacity. 
\citet{looper2008} calculated a small tangential velocity for 2M1821, which suggests a young age.
Therefore, 2M1821 could have  
low surface gravity \citep{gagne2014}, which may explain the large condensate opacity high in the atmosphere \citep{marley2012}. 
Our observations are consistent with such a scenario. 

\begin{figure}[ht]
  \begin{center}
    \includegraphics[scale=0.65,angle=0]{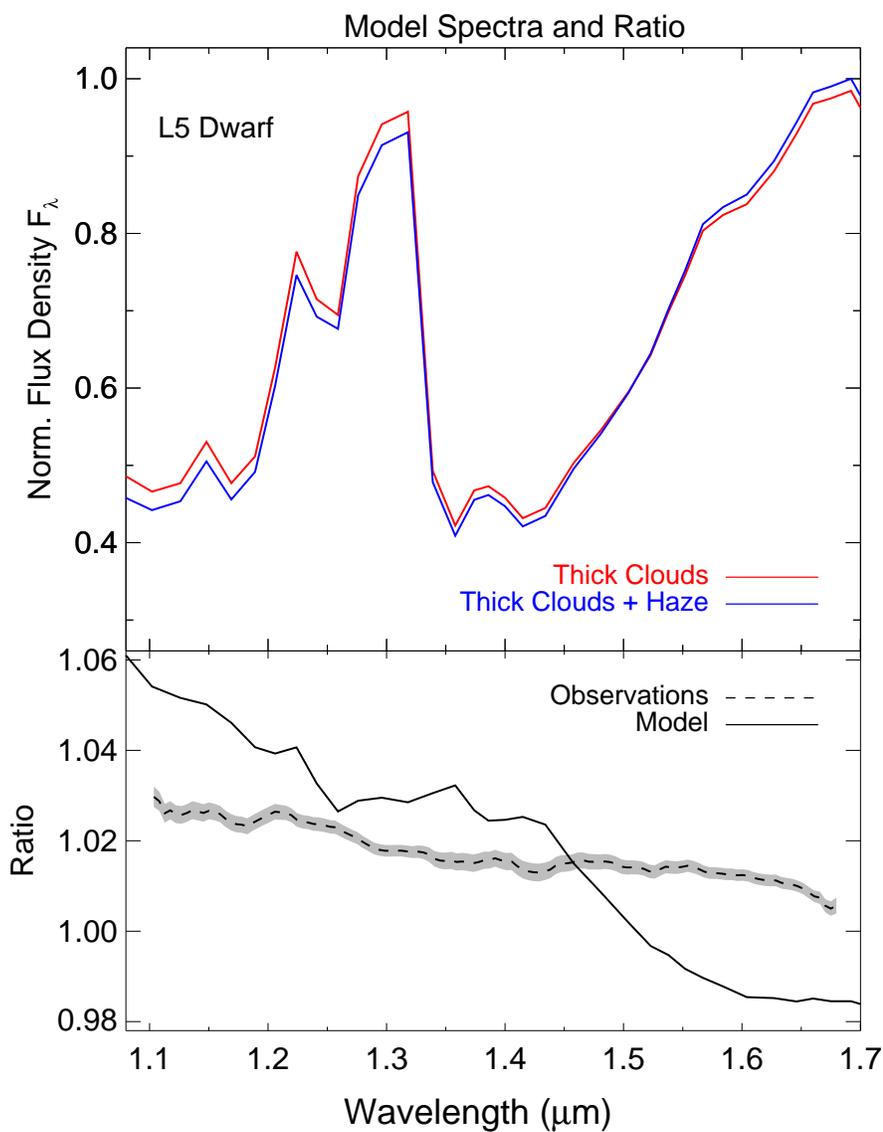}
       \caption{ Model spectra and their ratios for an L5 dwarf based on the models of \citet{saumon2008}. 
                 The observed spectral ratio (\emph{dashed line}) and associated uncertainties 
                 (\emph{gray band}) of the L5 dwarf 2M1821 are also shown for comparison.
                With a haze layer high in the atmosphere, models of an L5 dwarf reproduce a qualitatively similar
                spectral flux ratio as observed with 2M1821, especially the similar variation amplitude 
                in the 1.4 $\mu$m water absorption band with the adjacent continuum. 
                }

           \label{modelplot}
              \end{center}
    \end{figure}

In an attempt to reproduce the spectral variability observed for 2M1821, we also explored a variety of model 
atmosphere cases \citep{saumon2008}. 
By interpolating model spectra \citep{saumon2008} of different cloud thicknesses, \citet{radigan2012} were able to 
predict the relative spectral variations of the early T dwarf 2M2139, including the low water-band amplitude.
For 2M1821, we tried standard cloud models of \citet{ackerman2001} with varying cloud thickness, 
but no combination/modification of existing cloud models could explain the wavelength dependence of 
the observed variability amplitude.
To investigate if a spatially varying high-altitude haze might explain the observed variability, 
we compared two similar models. Both were for $T_{\rm eff}=1800\,\rm K$ and $\log g = 5$ 
with $f_{\rm sed} = 2$. We compared the emergent flux computed for a standard model with a model that additionally 
has a high, thin haze layer consisting of forsterite grains added to the top of the atmosphere. 
The haze particle size was set to have a single radius of $0.1\,\rm \mu m$ and was confined to pressures 
less than 50 mbar. 
The column geometric optical thickness of this haze (the total $\tau$ of the haze were it composed of 
perfectly scattering particles of the same size) is 0.7.  These haze particles are of
smaller size than the dust particles in the condensate clouds and have much longer settling timescales. 
Figure~\ref{modelplot} shows 
the ratio of the emergent flux of the two models in comparison with the observed spectral ratio, and we were able to produce 
the similar variability in and out of the water band with the haze model, further supporting the idea of dust particles 
residing high in the atmosphere of 2M1821.

While the shape of the spectral flux ratio is reproduced, we did not find a model flux ratio 
that is everywhere above unity as observed. The reason for this is that, when comparing two similar models, 
one with the high haze layer and one without, the model without the haze layer tends to always be cooler towards 
the top of the atmosphere. Thus the ratio of the two model spectra typically shows a flux excess (ratio $>1$) 
inside $J$ and $H$ bands, since the haze-free model allows more flux to escape within the window regions, which 
better sample the deeper and hotter parts of the atmosphere, 
but results in a flux deficit (ratio $<1$) inside of the water band as the less cloudy model is cooler. 
A more comprehensive model-fitting scheme will help better match the observed spectral ratio.




\section{CONCLUSION}
We have studied the near-IR spectral variablity of the L5 dwarfs 2M1821 and 2M1507 for the first time and found that 
the variablity of the 1.4 $\mu$m water band can probe the height of the cloud covers in the atmospheres. 
The weak wavelength dependence of spectral variations observed on the L5 dwarfs indicates that 
the dust grains giving rise to the flux variability likely reside at high altitudes, and this also fits 
general model predictions. We have found that the relative amplitude of flux variability out of the 1.4 $\mu$m water band 
compared with that in the water band displays an increasing trend from L5 dwarfs toward early T dwarfs.
Additional observations of objects with a range of spectral types are required to further confirm this trend.

With a limited sample of a few objects, we have demonstrated that rotational 
spectral mapping can be used to probe different atmospheric depths with different spectral features. 
When more advanced observing facilities such as the \emph{James Webb Space Telescope} become available, 
this technique can be applied to other atomic and molecular species and diagnose the atmospheric 
structures of brown dwarfs and directly imaged exoplanets in both horizontal and vertical directions. 
The high-altitude haze layers seen in brown dwarf atmospheres echo those found in exoplanetary atmospheres, 
further emphasizing the similarities in the properties of these ultracool atmospheres. Our observations 
open the possibilities of detailed comparative studies to understand the haze properties and behavior 
in both brown dwarf and exoplanetary atmospheres.


\acknowledgments 

This work is part of the Spitzer Cycle-9 Exploration Program Extrasolar Storms. This work is based in part 
on observations made with the Spitzer Space Telescope, which is operated by the Jet Propulsion Laboratory, 
California Institute of Technology under a contract with NASA. Support for this work was provided by NASA 
through an award issued by JPL/Caltech. Support for \emph{HST} GO programs 13176 and 13280.06-A was provided 
by NASA through a grant from the Space Telescope Science Institute, which is operated by the Association 
of Universities for Research in Astronomy, Inc., under NASA contract NAS5-26555. We acknowledge the outstanding 
help of Patricia Royle (STScI) and the Spitzer Science Center staff, especially Nancy Silbermann, for 
coordinating the HST and Spitzer observations. E.B. is supported by the Swiss National Science Foundation (SNSF).

\bibliographystyle{apj}





\clearpage

\begin{deluxetable}{ccccccccc}
\rotate
  \tablewidth{0pt}
  \tablecaption{ Journal of Observations. \label{obs}}
\tablecolumns{9}
\tabletypesize{\scriptsize}
\tablehead{
\colhead{Target} &  \colhead{Full Name} & \colhead{Spectral} &\colhead{\emph{J} } & \colhead{Date} & \colhead{Exposure} &\colhead{ $N_{\rm{exp}}$} &\colhead{No. of Orbits   } & Reference \\  
                &                      & \colhead{Type}     & \colhead{Mag.} &                        & \colhead{Time (s)} &\colhead{ per Orbit}      &\colhead{ $\times$ Visits}  &  }

\startdata
2M1507           & 2MASS J15074769-1627386  & L5             &  12.83 & 2013 Apr 30 \& May 12          & 67.30           & 30       & 4 $\times$ 2      & \citet{reid2000}  \\
2M1821           & 2MASS J18212815+1414010  & L5             &  13.43 & 2013 Jun 09 \& Jun 27          & 112.00          & 19       & 3 $\times$ 2      & \citet{looper2008} \\
SIMP0136         & 2MASS J01365662+0933473  & T2.5           &  13.45 & 2013 Sep 28 \& Oct 07          & 112.00          & 19       & 4 $\times$ 2      & \citet{artigau2006} \\


\enddata

\end{deluxetable}


\end{document}